\documentclass[twocolumn,preprintnumbers,showpacs,nofootinbib,superscriptaddress,leqn]{revtex4}
\usepackage{dcolumn}
\usepackage{amsmath,amssymb,setspace,graphicx,subfigure,epsfig}
\usepackage{amsfonts}
\usepackage{latexsym}
\usepackage{epsfig}
\usepackage{color}

\newcommand{\dis}[1]{\begin{equation}\begin{split}#1\end{split}\end{equation}}

\newcommand{\be}{\begin{equation}}
\newcommand{\ee}{\end{equation}}

\newcommand{\mplanck}{{M_P}}
\newcommand{\tev}{\,\textrm{TeV}}
\newcommand{\gev}{\,\textrm{GeV}}

\newcommand{\tadec}{T_{\tilde a\rm -dcp}}
\newcommand{\tarad}{T_{\tilde a\rm =rad}}
\newcommand{\treh}{T_{\rm R}}

\newcommand{\td}{T_{\rm D}}

\newcommand{\axino}{{\tilde{a}}}
\newcommand{\gravitino}{{\tilde{G}}}
\newcommand{\gluino}{{\tilde{g}}}

\begin{document}

\title{Neutralino dark matter from heavy axino decay}

\author{Ki-Young Choi}
\email{kiyoung.choi@uam.es}
\affiliation{Departamento de F\'{\i}sica Te\'{o}rica C-XI,
        Universidad Aut\'{o}noma de Madrid, Cantoblanco,
        28049 Madrid, Spain}
\affiliation{Instituto de F\'{\i}sica Te\'{o}rica UAM/CSIC,
        Universidad Aut\'{o}noma de Madrid, Cantoblanco,
        28049 Madrid, Spain}

\author{Jihn E. Kim}
\email{jekim@phyp.snu.ac.kr}
\affiliation{ Department of Physics and Astronomy, Seoul National University, Seoul 151-747, Korea}

\author{Hyun Min Lee}
\email{hmlee@andrew.cmu.edu}
\affiliation{Department of Physics, Carnegie Mellon University, 5000 Forbes Avenue, Pittsburgh, PA 15213, USA.}

\author{Osamu Seto}
\email{osamu.seto@uam.es}
\affiliation{Instituto de F\'{\i}sica Te\'{o}rica UAM/CSIC,
        Universidad Aut\'{o}noma de Madrid, Cantoblanco,
        28049 Madrid, Spain}


\begin{abstract}
We consider cosmological consequences of a heavy axino, decaying to the neutralino in R-parity conserving models.
The importance and influence of the axino decay
on the resultant abundance of neutralino dark matter
depends on the lifetime and the energy density of axino.
For a high reheating temperature after inflation,
copiously produced axinos dominate the energy density of the universe and its decay produces a large amount of entropy.
As a bonus, we obtain that the upper bound on the reheating temperature after inflation  via gravitino decay can be moderated, because the entropy production by the axino decay more or less dilutes the gravitinos.
\end{abstract}

\preprint{IFT-UAM/CSIC -08-01}

\maketitle

\section{Axino}

Neutralino, if it is the lightest supersymmetric particle (LSP) in R-parity conserving models, is a natural candidate for dark matter. Because of the TeV scale sparticle interactions, the thermal history of neutralinos allows the neutralino dark matter possibility. But, imposing a solution of the strong CP problem, the thermal history involves contributions from the additional sector.

The strong CP problem is naturally solved by introducing a very light axion $a$. Most probably, it appears when the Peccei-Quinn (PQ) symmetry is broken at a scale of $f_a$.
Below the PQ scale, the effective axion interaction with gluons is
\dis{
{\cal L} =\frac{g_s^2}{32\pi^2 f_a}a F \tilde{F},
}
where $g_s$ is the strong coupling constant~\cite{Review}.
The PQ scale
is constrained by the astrophysical and cosmological
considerations in the narrow window
$10^{10} \gev \lesssim f_a \lesssim 10^{12} \gev$~\cite{fabound}.

TeV scale supersymmetry (SUSY) suggests axino $\tilde a$, the superpartner of axion, around the electroweak scale in the gravity mediation scenario. Here, we consider the effects of {\it heavy} axinos in cosmology. The axino cosmology depends crucially on the axino decoupling temperature \cite{Rajagopal:1990yx},
\dis{
\tadec =10^{11}\gev \left(\frac{f_a}{10^{12}\gev} \right)^2\left(\frac{0.1}{\alpha_s} \right)^3,\label{T_dec}
}
where $\alpha_{s}=g_s^2/4\pi$.

The axion supermultiplet includes axion, saxion (the scalar partner) and axino. Both saxion and axino masses are split from the almost vanishing axion mass if SUSY is broken. The precise value of the axino mass depends on the model, specified by the SUSY breaking sector and the mediation sector to the axion supermultiplet~\cite{susyreview}. In principle, the axion supermultiplet is independent from the observable sector in which case we may take the axino mass as a free parameter of order from keV to a value much larger than the gravitino mass \cite{axinomass,ChunLukas}. Light axinos can be a dark matter (DM) candidate, which has been studied extensively \cite{axinoDM,Roszkowski:2006kw,Asaka:2000ew}. Heavy axinos, however, cannot be the LSP and can decay to the LSP plus light particles. This heavy axino decay to neutralino was considered in the literature \cite{ChunLukas} where the neutralino relic density was not considered seriously. Some considered the axino as the next LSP decaying to the gravitino LSP in the gauge mediated SUSY breaking scenario \cite{Chun:1993vz}. Recently, supersymmetric axion models were studied with an emphasis on saxion \cite{Kim:1992eu}, where the heavy axino possibility was also considered briefly \cite{Kawasaki:2007mk}.

In this paper, we present a more or less complete cosmological analysis of a {\it heavy} axino with mass in the TeV region so that it is heavier than the LSP neutralino. Compared with the saxion study, the heavy axino study probes SUSY directly because the axino carries the odd R-charge.

If kinematically allowed, a heavy axino decays predominantly to a gluino and a gluon and the gluino subsequently decays, finally producing the neutralino LSP. However, if axino is lighter than gluino, this axino to gluino decay is forbidden and the axino predominantly decays to a neutralino and a photon.
The neutralinos from axino decay can annihilate in the cosmos if the neutralino number density, $n$, is large enough so that $n \langle \sigma_{ann} v\rangle > H$ where $\sigma_{ann}$ is the annihilation cross section. In this case, the neutralino abundance is modified, which is obtained by solving the Boltzmann equation. For this to be compatible with the observed DM density, we find the required
thermally-averaged cross section of neutralino,
$ \langle \sigma_{ann} v\rangle $, to be around $10^{-8}\gev^{-2}$.
The Higgsino-like neutralino can give this kind of large cross section.

The thermally produced ${\cal O}(100 \gev)$ gravitinos after inflation induce severe problems on the light element abundances, which restricts the reheating temperature to $\treh< 10^{6-7} \gev$. We find that the entropy production from the heavy axino decay can dilute the primordial gravitinos.
For axino mass smaller than gluino mass and for a low PQ scale, $f_a\sim 10^{10}\gev$, we find as shown below that there is no gravitino problem for the reheating temperature up to the axino decoupling temperature $\tadec$.

A heavy axino leads to different physical consequences depending on its mass being greater or smaller than the gluino mass $m_\gluino$. If axino is heavier than gluino, it decays dominantly to a gluino plus a gluon.  If axino is lighter than gluino, it decays dominantly to a $b$-ino-like neutralino and a
photon and, if kinematically allowed, to a neutralino and a $Z$-boson. Thus, the axino decay width is given by,

\noindent (i) $m_{\tilde a}>m_\gluino$
\dis{
\Gamma(\axino \rightarrow \gluino + g)=\frac{8\alpha_{s}^2}{128\pi^3}
\frac{m_\axino^3}{f_a^2}
\left(1-\frac{m_\gluino^2}{m_\axino^2} \right)^{3},
\label{Gamma_axino}
}
\noindent (ii) $m_\chi<m_{\tilde a}<m_\gluino$:
\begin{equation}
\Gamma(\axino \rightarrow \chi_i + \gamma)=
 \frac{\alpha_{em}^2C^2_{a\chi_i\gamma}}{128\pi^3}
\frac{m_\axino^3}{f_a^2}
\left(1-\frac{m_{\chi_i}^2}{m_\axino^2} \right)^{3},
\end{equation}
with
$C_{a\chi_i\gamma}=(C_{a Y Y}/\cos\theta_W)Z_{\chi_i B}$~\cite{axinoDM},
where $Z_{\chi_i B}$ is the $b$-ino fraction of the $i$-th neutralino and  $\theta_W$ is a Weinberg mixing angle. Hereafter we will use $C_{a Y Y}=Z_{\chi_i B}=1$ for simplicity.
For the case (ii), as one can see,
the lifetime can be easily longer than $0.1$ second.
If the lifetime is shorter than about $0.1$ second,
the axino decay does not harm the standard big bang nucleosynthesis (BBN). Note that for Eq.~(\ref{Gamma_axino}) the axino lifetime is about $3.3\times 10^{-7} \textrm{s}
\left({\alpha_{s}}/{0.1} \right)^{-2}
\left({f_a}/{10^{11} \gev} \right)^2
\left({m_\axino}/{1 \tev} \right)^{-3}$.

Right after the axino decay, the temperature of radiation $\td= { \sqrt{\Gamma_\axino M_P}}/{\left(\pi^2g_*/90\right)^{1/4}} $ is given by
\dis{
\td = 1.4\ {\rm GeV}
 \left(\frac{70}{g_*} \right)^{1/4}
 \left(\frac{3 \times 10^{-7} {\rm sec}}
 {\tau_\axino}\right)^{1/2},
\label{TD}
}
where $M_P$ is the reduced Planck mass.
For the case that axinos dominate the energy density of the universe before they decay, we can regard $\td$ as
the second reheating temperature due to the axino decay.
This temperature is marginal when we discuss its effect on neutralino DM as we discuss below. This is because, for the axino lifetime of $10^{-7}$ second, this temperature is comparable to the typical neutralino freeze-out temperature $T_{fr} \approx m_{\chi}/25$ \cite{KolbTurner}.
If $\td > T_{fr}$ or equivalently $\tau_{\tilde{a}}<{\cal O}(10^{-7})$ second, the axino decay has no effect on the neutralino DM abundance, because the neutralino produced by the axino decay also could reach the thermal equilibrium.
From Eq.~(\ref{Gamma_axino}), we can see that this would be the case for a heavier axino or for a lower PQ scale.

The axino abundance depends on the thermal history
of the early universe after inflation. Another relevant temperature we introduce is the reheating temperature after inflation $\treh$. So, the temperatures we introduce are
\dis{
&\tadec={\rm axino\ decoupling\ temperature}\\
&\treh={\rm reheating\ temperature\ after\ inflation}\\
&T_{fr}={\rm neutralino\ freeze-out\ temperature}\\
&\tarad={\rm axino-radiation\ equality\ temperature}\\
&\td = {\rm radiation\ temperature\ right\ after\ \tilde {\it a}\ decay}.
\label{Temps}
}

For $\treh>\tadec$, axinos were in the thermal equilibrium and the axino number density is comparable to the photon number density. Below the axino decoupling temperature, the axino number in the comoving volume is conserved
if there is no new physics below $\tadec$; then
the axino abundance is given by
\dis{
Y_\axino=\left.\frac{n_\axino}{s}\right|_{\tadec}
=\frac{135\zeta(3)}{8\pi^4}
\frac{g_\axino}{g_*(\tadec)},\quad \textrm{for}\quad \treh>\tadec,
}
where $\zeta(3)\simeq1.202$, $g_\axino$ is the degrees of freedom of axino and $s=\frac{2\pi^2}{45}g_{*s}T^3$ is the entropy density.

For $\treh<\tadec$, axinos could not be in thermal
equilibrium after inflation. Nevertheless, they are regenerated by thermal scattering and by decays of gluinos, squarks and neutralinos in the thermal plasma.
For a high enough reheating temperature, $T\gg 10^{4}\gev$,
the hard thermal loop approximation is good enough and the axino abundance can be approximated as \cite{Brandenburg:2004du}
\dis{
Y_{\tilde{a} } = 2.0\times 10^{-7}g_s^6 \ln \left( \frac{1.108}{g_s}\right)\left(\frac{10^{11} \gev}{f_a} \right)^2\left( \frac{T_R}{10^4\gev}\right).
}
We can see that
the axino abundance might be large in terms of $Y_{\tilde{a}}$ going up to ${\cal O}(0.002)$, strongly depending on $\treh$ and $f_a$. Hence, if many axinos are produced after inflation, axinos might dominate the energy density of the universe before they decay.

Let us first examine the condition for axino domination.
For $T < m_{\tilde{a}}$, axino becomes nonrelativistic.
When axinos are nonrelativistic, the energy densities of radiation and axinos  are given by
\begin{eqnarray}
\rho_R = \frac{\pi^2 g_*}{30} T^4
,\quad \left.\rho_{\tilde{a}} \right|_{T<m_{\tilde{a}}}= m_{\tilde{a}} Y_{\tilde{a}} s  \label{axino:Tma},
\end{eqnarray}
respectively. From Eq.~(\ref{axino:Tma}), we find
the axino-radiation equality temperature, $\tarad$,
\begin{equation}
\tarad\equiv \frac{4}{3} m_{\tilde{a}} Y_{\tilde{a}}> \td.
\label{DominationCondition}
\end{equation}
If this equality occurs before axino decays as shown as the inequality, then axino can dominate the universe.
On the other hand, for $ \frac{4}{3} m_{\tilde{a}} Y_{\tilde{a}} < \td$, axinos never dominate the energy density of the universe before they decay. In this regard, we find the lowest reheating temperature, $\treh^{min}$, above which axinos can dominate the universe before they decay, by solving the equality
$
 \frac{4}{3} m_{\tilde{a}} Y_{\tilde{a}}(\treh^{min}) = \td.
$

In case axinos dominate the universe, the entropy production by axino decay dilutes the previously existing number densities. The ratio of the entropy per comoving volume before and after the axino decay is \cite{KolbTurner}
\dis{
r\equiv\frac{S_{f}}{S_{0}}\simeq \frac{4 m_{\tilde{a}} Y_{\tilde{a}}}{3 \td} ,
\label{EntropyProduction}
}
for $ \tarad > \td $. Thus, from Eqs.~(\ref{TD}) and (\ref{EntropyProduction}), the entropy ratio can be
\dis{
r\equiv\frac{S_{f}}{S_{0}} \simeq 1.3
\left(\frac{m_\axino}{1\tev}\right)
\left(\frac{2\gev}{\td}\right)
\left(\frac{Y_\axino}{0.002}\right)
\label{entropyratio}.
}
This is shown in Figs.~\ref{fig_fa10} and~\ref{fig_fa12} as magenta lines. Above the solid line denoted as $r=1$, axinos can dominate the universe and can produce additional entropy.

Similarly, in case that an axino can decay only into a neutralino and a photon [Case (ii)], $T_{\rm D}$ looks to be able to become as small as of several MeV and $S_{f}/S_{0}$ might be as large as ${\cal O}(10^2-10^3)$.

\section{Relic density of neutralino}

Neutralino, which was in the thermal equilibrium in the early universe, decouples and freezes out when the annihilation rate becomes smaller than the Hubble parameter.
The freeze-out temperature $T_{fr}$ is normally given by $m_\chi/25$, e.g. $4 \gev$ for $100 \gev $ neutralino. As mentioned above, the neutralino relic density is not affected by axino for $\td > T_{fr}$.

Hence, we consider  $\td < T_{fr}$ below and the following arguments are valid only for this case. Another crucial issue in this case is whether the produced neutralinos from the axino decay would annihilate again or not. If the neutralino number density from axino decay is too large, i.e. for
\begin{equation}
\left. Y_{\chi}\right|_{T=\td} > \left(\frac{90}{\pi^2 g_*}\right)^{1/2}
\frac{1}{4\langle\sigma v\rangle}\frac{1}{M_P \td},
\label{AnnihilationCondition}
\end{equation}
the produced neutralinos would annihilate more.

\subsection{Case of axino domination}

For $\td < T_{fr}$, let us consider the cases of axino domination. First, let us consider the case that axinos would dominate the universe but the produced neutralino cannot annihilate. From Eqs.~(\ref{DominationCondition})
and (\ref{AnnihilationCondition}), this occurs if
\begin{eqnarray}
\langle\sigma v\rangle
 < \frac{\left(90/\pi^2 g_*\right)^{1/2}}{4 M_P \td Y_\chi(\td)}
 \simeq \frac{m_{\tilde{a}}}{3 M_P \td^2}\left(\frac{90}{\pi^2 g_*}\right)^{1/2}
\label{DominationNoAnnihilation}
\end{eqnarray}
is satisfied. The equality in (\ref{DominationNoAnnihilation}) is estimated using
Eq.~(\ref{entropyratio}) assuming that the neutralino abundance from axino decay is much bigger than
that from thermal freeze-out. This condition is not likely to be satisfied for the range of our $m_\axino$ and $\td$.

This leads us to the case that Eq.~(\ref{DominationNoAnnihilation}) is not satisfied, i.e.
neutralinos produced (from axino decay and thermal freeze-out) annihilate. The final abundance is obtained solving the Boltzmann equation,
\begin{equation}
\frac{d n_\chi}{d t}+3Hn_\chi = - \langle\sigma_{ann} v_{rel} \rangle n_\chi^2,
\label{BoltzmannEq}
\end{equation}
where $\sigma_{ann}$ is the annihilation cross section of two neutralinos, $v_{rel}$ is their relative velocity and $\langle  \ldots \rangle$ is the thermal average.
With $Y_\chi\equiv n_\chi/s$, the evolution equation is
\dis{
\frac{d Y_\chi}{d t}=
-\langle\sigma_{ann} v_{rel} \rangle Y_\chi^2 s,
}
where $s$ is the entropy density, and $s\propto T^{3}\propto t^{-3/2}$ in the radiation dominated era.

Applying the sudden decay approximation,
after the decay of axinos, the radiation and neutralino densities can be easily estimated. The radiation dominates the universe soon and the neutralinos start to annihilate.
We can solve this evolution equation from the time after axino decay, $\td$, to later times. The final abundance $Y_\chi$ at temperature $T$ after neutralino annihilation
can be expressed as
\dis{
Y_\chi^{-1}(T)=& Y^{-1}_\chi(\td)- \langle\sigma_{ann} v_{rel} \rangle
\left( \frac{s}{H}-  \frac{s(\td)}{H(\td)}\right) \\
\simeq&  Y^{-1}_\chi(\td)+\frac{ \langle\sigma_{ann} v_{rel} \rangle
s(\td)}{H(\td)}.
\label{finalY:neutralino}
}
Here, $Y_\chi(\td)$ is the sum of neutralino densities from thermal freeze-out and from the axino decay, right after axino decay but before neutralino
annihilation,
\begin{equation}
 Y_\chi(\td) = Y_\chi^{fr}+ Y_\chi^{decay}.
\label{initialY:neutralino}
\end{equation}
When axino dominates before the neutralino freeze-out, $\tarad >T_{fr}$, then the freezed-out neutralino relic density, which is different from the standard one in the radiation dominated universe, can be obtained as
\begin{eqnarray}
&&Y_\chi^{fr} =  \frac{(90/\pi^2 g_{fr})^{1/2}}
 {4 \langle\sigma v \rangle M_P T_{fr}^{aD}}
  \frac{3}{2} \left(\frac{g_D}{g_{fr}}\right)^{1/2}
  \left(\frac{\td}{T_{fr}^{aD}}\right)^3, \label{FrzOut:neutralino}
\end{eqnarray}
where $g_D=g_*(\td)$, $g_{fr}=g_*(T_{fr}^{aD})$
and $T_{fr}^{aD}$ is the freeze-out temperature of the neutralino in the nonrelativistic axino matter dominated universe.
Otherwise, i.e. for  $\tarad<T_{fr}$ we obtain
\begin{eqnarray}
 Y_\chi^{fr} =  \frac{(90/\pi^2 g_*(T_{fr}))^{1/2}}
 {4 \langle\sigma v \rangle M_P T_{fr}}.\label{Yfr_rad}
\end{eqnarray}

\begin{figure}[t]
  \includegraphics[width=9cm]{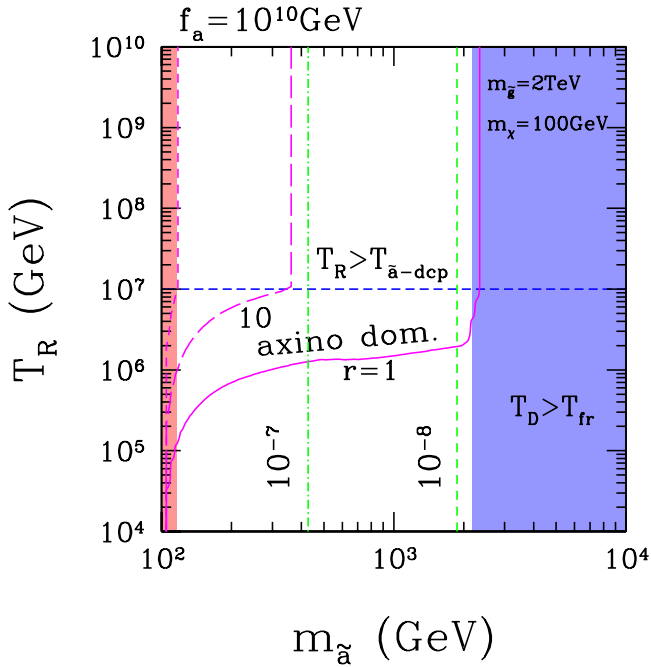}\\
  \caption{The $\treh$ vs. $m_\axino$ plot for $f_a=10^{10} \gev$. The region $\treh>\tadec$ is above the dashed blue line (horizontal). The axino lifetime greater than $0.1$ sec is denoted by the red shaded region in the left side. The blue shaded region in the right side is where axino decays before neutralino decouples ($\td > T_{fr}$). The magenta lines (horizontal) are the contours of the entropy increase, $r\equiv S_f/S_0$. Above $r=1$ lines axinos dominate the universe before they decay. The green lines (vertical) denote the $\langle \sigma_{ann} v_{rel}\rangle$ in units of ${\rm GeV}^{-2}$ which are used to give the right amount of neutralino relic density by Eq.~(\ref{neutralinorelicdensity}). We use neutralino and gluino masses as $m_\chi=100\gev$ and $m_\gluino = 2 \tev$. For a heavier neutralino mass, the green lines move to the right. }\label{fig_fa10}
\end{figure}
\begin{figure}
  \includegraphics[width=9cm]{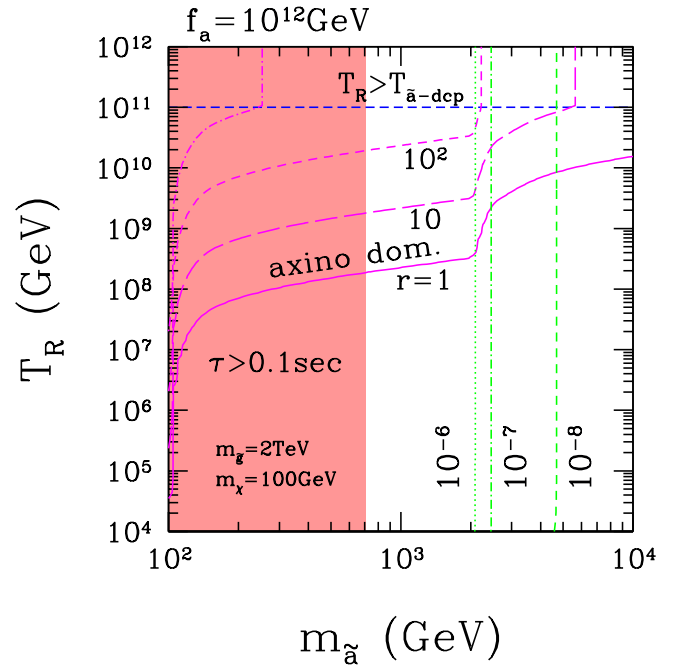}\\
  \caption{The same as Fig. \ref{fig_fa10} but with $f_a=10^{12} \gev$.}\label{fig_fa12}
\end{figure}

The relic density of neutralinos from the axino decay can be given by the axino abundance suppressed by the entropy production,
\begin{eqnarray}
 Y_\chi^{decay} = Y_{\tilde{a}}\times \frac{S_0}{S_f} ,
 \label{NTP:axino}
\end{eqnarray}
where $Y_{\tilde{a}}$ is the axino abundance just before axino decay.

When the number density of neutralinos from the axino's sudden decay is much larger, so that $n_\chi(\td) \langle\sigma_{ann}v_{rel}\rangle \gg H(\td)$,
then we obtain the result of~\cite{Nakamura:2007wr}.
In this limit, the relic density can be approximated and simplified as
\dis{
\Omega_\chi h^2 \simeq  &
0.14 \left(\frac{90}{\pi^2 g_*(\td)} \right)^{1/2}
\left(\frac{m_\chi}{100 \gev} \right)\\
&\times
\left( \frac{10^{-8}\gev^{-2}}{ \langle\sigma_{ann} v_{rel} \rangle }\right)
\left(\frac{2 \gev}{\td} \right),\label{neutralinorelicdensity}
}
where we normalized $\td$ for an axino decay to gluino and gluon. The relic density of neutralino is proportional to the neutralino mass which is different from the standard one where $\Omega_\chi h^2 \propto x_f\equiv m_\chi/T_{fr} \sim25$.

\subsection{Case of axino nondomination}

In the case that axinos never dominate the universe,
$\tarad < \td$, the resultant neutralino LSP abundance is given by Eqs.~(\ref{finalY:neutralino}) and (\ref{initialY:neutralino}),
but with  $Y_\chi^{fr}$ given by eq.~(\ref{Yfr_rad}) and
$Y_\chi^{decay} = Y_{\tilde{a}}$,
in which $T_{fr}$ is the freeze out temperature of the neutralino in the radiation dominated universe, instead of Eqs.~(\ref{FrzOut:neutralino}) and (\ref{NTP:axino}).

If Eq.~(\ref{AnnihilationCondition}) is satisfied, as in the previous case, we obtain the same final neutralino relic abundance as that given in Eq.~(\ref{neutralinorelicdensity}). This is evident if we think that the final neutralino relic density
after re-annihilation is determined from the Hubble parameter at the time of axino decay.

On the other hand, if Eq.~(\ref{AnnihilationCondition}) is not satisfied, we simply obtain
\begin{eqnarray}
 Y_\chi =  \frac{(90/\pi^2 g_*(T_{fr}))^{1/2}}
 {4 \langle\sigma v \rangle M_P T_{fr}} + Y_{\tilde{a}} .
\end{eqnarray}
This is possible when the axino abundance is too small, which is the case for low reheating temperature $\treh\lesssim \mathcal{O}(10^2 \gev)$. For a large annihilation cross section, the first term becomes negligible and the second term is dominant. In this case, the neutralino number density is given by thermally produced axino's. Hence, in this case also, we may have a chance to measure the reheating temperature as in the axino DM case, as recently pointed out
by Choi {\it et. al.} \cite{Choi:2007rh}.

\begin{figure}
  \includegraphics[width=9cm]{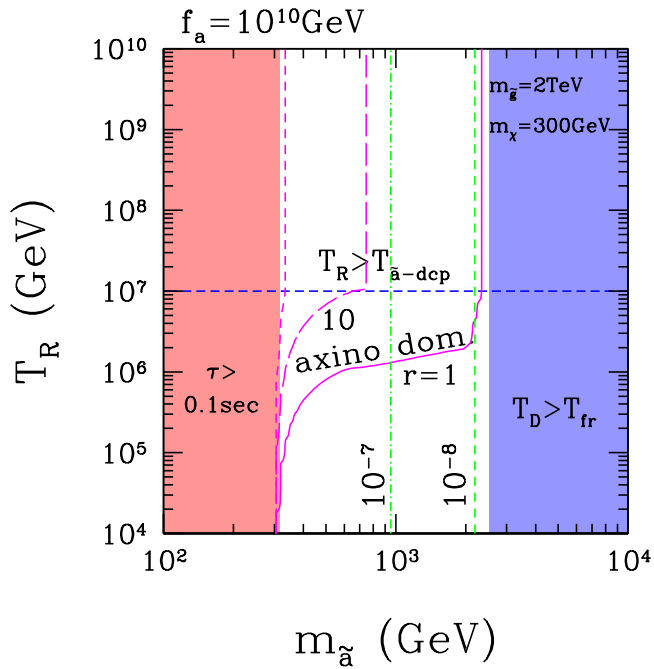}\\
  \caption{
The same as Fig. \ref{fig_fa10} but with $m_\chi=300 \gev$.}\label{fig_fa10_mchi}
\end{figure}

\begin{figure}
  \includegraphics[width=9cm]{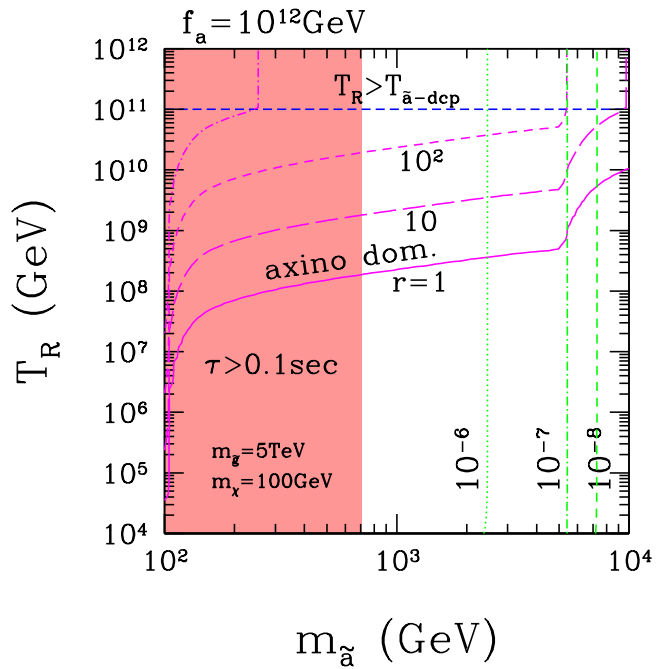}\\
  \caption{
The same as Fig. \ref{fig_fa12} but with $m_{\tilde{g}}=5 \tev$.}\label{fig_fa12_mgluino}
\end{figure}

In Figs.~\ref{fig_fa10} and~\ref{fig_fa12}, we show the contour lines (green lines) of $ \langle\sigma v \rangle$ which gives the neutralino relic density suitable for the DM in the universe. As can be seen in the figures, the high annihilation cross section of neutralinos of order $\langle\sigma v \rangle \gtrsim 10^{-8} \gev$ is necessary.
Here, we used the neutralino mass of $100 \gev$. If we increased the neutralino mass, then the  $\langle\sigma v \rangle$ increases accordingly since neutralino relic density is proportional to the neutralino mass as shown in Eq.~(\ref{neutralinorelicdensity}), which is plotted in Fig.~\ref{fig_fa10_mchi}
with $m_\chi=300 \gev$.
For completeness we show the plots with different gluino mass in Fig.~\ref{fig_fa12_mgluino}, where we used $m_\gluino= 5 \tev$.


\section{Softening the Gravitino Problem}
In the early universe, gravitinos are produced after reheating. It is known that the gravitino abundance is proportional to the reheating temperature $\treh$. If the gravitino is not the LSP then it can decay. Since the interaction of gravitino is suppressed by the Planck mass,  the lifetime is much longer than that of axino, usually falling in the BBN or post-BBN era.
The decay products of gravitino can change the abundances of light elements and gives the severe constraint on the reheating temperature, $\treh < 10^{6-7}\gev$ for $m_\gravitino\sim \tev$ \cite{BBNgrav,Kohri:2005wn}.

For a high reheating temperature in our scenario,
the axino dominates the universe and excessive entropy is produced as shown in Eq.~(\ref{entropyratio}), which may soften the gravitino problem. This entropy production is proportional to the reheating temperature, which is of the same form as the gravitino production. Therefore, as $\treh$ increases, the increased number of gravitinos is diluted by the increased entropy from the axino decay, which renders the gravitino abundance independent of the reheating temperature after axino enters into the axino dominated phase in the universe. This independence is true only when $\treh$ is smaller than the axino decoupling temperature $\tadec$.

The abundance of gravitinos $Y_{\gravitino}$
after axino decay can be expressed as
\dis{
Y_{\gravitino}=Y^{th}_\gravitino(\treh)
\left(\frac{S_{f}}{S_{0}}\right)^{-1}=
Y^{th}_\gravitino(\treh^{min}),
}
where $Y^{th}_\gravitino(\treh)$ is the thermally produced gravitino abundance after reheating. $\treh^{min}$ is the lowest reheating temperature above which the axino dominates: $r=1$, denoted by the solid magenta line in the figures. This is valid for $\treh^{min} < \treh < \tadec$.

So for $f_a=10^{10} \gev$, this line is almost $\treh^{min}=10^6 \gev$, where the gravitino problem is almost solved anyway. However, for $f_a=10^{12} \gev$,  $\treh^{min}=\textrm{around}\ 10^8 \gev$, the gravitino problem  still exists, though it is tolerated in our heavy axino scenario.

\section{Conclusion}

We have discussed the heavy axino possibility so that the heavy axino decay leads to a reasonable DM density of the LSP neutralino and enough radiation to dilute the gravitinos. For this scenario to be realized, we need a large annihilation cross section of neutralinos $\langle\sigma v \rangle$.
Such a neutralino is Higgsino-like, in which case the cross section with nuclei for the direct DM search can be as large as $\sigma_{SI} \simeq 10^{-(7-8)}$ pb, which can be probed in future DM search experiments.

\vskip 0.5cm


\acknowledgments
K.-Y.C. is supported by the Ministerio de Educacion y Ciencia of Spain under Proyecto Nacional FPA2006-05423 and by the Comunidad de Madrid under Proyecto HEPHACOS, Ayudas de I+D S-0505/ESP-0346, J.E.K. is supported in part by the KRF Grants, No. R14-2003-012-01001-0 and No. KRF-2005-084-C00001, H.M.L. is supported by  DOE Contracts DOE-ER-40682-143
and DEAC02-6CH03000, and O.S. is supported by the MEC project (No.FPA 2004-02015) and the Comunidad de Madrid project HEPHACOS (No.~P-ESP-00346).
K.-Y.C. and O.S. would like to thank the European Network of Theoretical 
Astroparticle Physics ILIAS/ENTApP under contract number 
RII3-CT-2004-506222 for financial support.
%

\appendix*
\section{Neutralino freeze-out density in decaying particle dominated cosmology}

In this appendix, we derive an approximate relic density of
thermal neutralinos decoupled in the decaying matter dominated universe, Eq.~(\ref{FrzOut:neutralino}) for completeness. Although one may find the same discussion in Ref. \cite{McDonald:1989jd}, some $g_*$ dependence is missing there and we will correct it. Just after the decoupling of neutralino, from the relevant Boltzmann equation (\ref{BoltzmannEq}), we obtain
\begin{equation}
\frac{1}{\left. a^3 n\right|_t}
 -\frac{1}{\left. a^3 n\right|_{t_{fr}}}
 \simeq \left.\frac{\langle \sigma v \rangle 2}{a^3 3H} \right|_{t=t_{fr}} ,
\end{equation}
for $t\gg t_{fr}$. In the matter dominated universe $t_{fr}=2/3H_{fr}$, and the freeze-out temperature is
$T_{fr}\simeq -m_\chi \ln\left[\frac{ 3\sqrt5 \langle \sigma v \rangle \mplanck m_\chi^{3/2}g_f^{1/2}T_f^{2}}{\pi^{5/2}
g_{fr}T_{fr}^{5/2}}\right]$.
Here, in the decaying matter dominated universe,
the entropy density per comoving volume is not conserved.
Hence, we used $a^3$ instead of the entropy density in the standard calculation.
After the reheating by the decaying matter is completed,
when the temperature is $T_f$, of course, we can use $Y$ as usual. We find
\begin{equation}
\left. Y\right|_{T=T_f} \simeq
\left(\frac{a({T_{fr})}}{a(T_f)}\right)^3 \frac{3H_{fr}}{2\langle \sigma v \rangle s(T_f) },
\end{equation}
after the decay. Recalling $g_*(T) T^4 \propto H$ and $a(T)^3\propto g_*(T)^2T^8$ in the decaying matter dominated universe, we obtain
\begin{eqnarray}
Y \simeq \left(\frac{g_*(T_f)}{g_{*S}(T_f)}\right)
 \left(\frac{90}{\pi^2 g_*(T_{fr})}\right)^{1/2}
 \frac{1}{4 \langle \sigma v \rangle M_P T_{fr}} \nonumber \\
\times \frac{3}{2}
\left(\frac{T_f}{T_{fr}} \right)^3 \left(\frac{g_*(T_f)}{g_*(T_{fr})}\right)^{1/2} .
\end{eqnarray}
%
In text, we used $g_*(T_f) = g_{*S}(T_f)$ and $T_f=\td$.


\end{document}